\documentclass[12pt]{article}
\usepackage{latexsym,cite}
\begin{document}
\makeatletter
\@addtoreset{equation}{section}
\renewcommand{\theequation}{\thesection.\arabic{equation}}
\makeatother

\newcommand{\lp}{\left}
\newcommand{\rp}{\right}
\newcommand{\be}{\begin{equation}}
\newcommand{\ee}{\end{equation}}
\newcommand{\<}{\langle}
\renewcommand{\>}{\rangle}
\newcommand{\reff}[1]{(\ref{#1})}
\newcommand{\pslash}{\hbox{$k_1$\kern-0.9em\raise-0.0ex\hbox{$/$}}\thinspace}
\newcommand{\qslash}{\hbox{$k_2$\kern-0.9em\raise-0.0ex\hbox{$/$}}\thinspace}
\newcommand{\kslash}{\hbox{$k$\kern-0.5em\raise-0.0ex\hbox{$/$}}\thinspace}

\title{The $\pi_0 \rightarrow \gamma \gamma$ decay and the chiral
  anomaly in the  quark-composites approach to QCD} 

\author{
  \\
  { Sergio Caracciolo }              \\
  {\small\it Scuola Normale Superiore and INFN -- Sezione di Pisa}  \\[-0.2cm]
  {\small\it I-56100 Pisa, ITALIA}          \\[-0.2cm]
  {\small Internet: {\tt sergio.caracciolo@sns.it}}   
  \\  \and
  { Fabrizio Palumbo~\thanks{This work has been partially
  supported by EEC under TMR contract ERB FMRX-CT96-0045}}\\
  {\small\it INFN -- Laboratori Nazionali di Frascati}  \\[-0.2cm]
  {\small\it P.~O.~Box 13, I-00044 Frascati, ITALIA}  \\[-0.2cm]
  {\small Internet: {\tt palumbof@lnf.infn.it}}
  }

\date{\today}
\maketitle
\thispagestyle{empty}

\begin{abstract}
We evaluate the $\pi_0 \rightarrow \gamma \gamma$ decay amplitude by
an effective  
action derived from QCD in the quark composites approach, getting the
standard value. 
We also verify that our effective action correctly reproduces the
chiral anomaly.  
\end{abstract}
\clearpage
\section{Introduction}

Unlike previous field theories, QCD does not contain the fields
which describe the particles observed in the experiments. 
This motivated the introduction of effective Lagrangians like the 
chiral Lagrangians, written in terms of the phenomenological fields
and based solely on general principles of invariance.

The chiral Lagrangians encode the spontaneous breaking of the chiral symmetry 
which is generally believed to occur in QCD and the old results 
of current algebra and PCAC~\cite{Wein,GL}. But while these descriptions
have been 
put on the safe ground of a consistent and predictive field theory,
the determination  
from QCD of the parameters appearing in them as well as the proof that
chiral symmetry
is spontaneously broken in this theory are still lacking. The foundations
of the chiral Lagrangians could, in principle, be completed by the use of 
numerical simulations in the lattice formulation of QCD,
but, of course, an  
analytical approach is desirable also for a better understanding and a
more clever  
way of using the numerical recipes. There have been, indeed, attempts
in this direction, but they have been so 
far restricted to the strong coupling region  in the gauge coupling
constant~\cite{Kawa,Saclay1,Saclay2,Rebbi1,Rebbi2}, which  is
unfortunately far from the interesting continuum 
limit.

It has also been proposed to extend QCD via the introduction of
extra degrees of  freedom~\cite{Brower:1994,Brower:1995}.   
The corresponding fields, even though carry the quantum numbers of the
chiral mesons, are supposed to decouple in the continuum limit, in
order to avoid double counting.

The idea behind the quark composites approach is that it should be
possible to recover the 
interactions of the phenomenological fields by a change of variables
in the partition 
function of QCD whereby the quark composites with the quantum numbers
of the phenomenological fields of interest are
assumed as new integration variables. The final goal is to unify the
description of the spectrum properties and scattering processes in
a framework consistent with the confinement of quarks.

For technical reasons this program can be realized in this form only for the
baryons~\cite{DeFr1,DeFr2}. For the mesons instead of a change of 
variables we make recourse to auxiliary fields~\cite{Cara}. This
latter procedure,  
however, should not be confused with the quoted use of additional
phenomenological fields in QCD. 

In practice, to implement our approach in a perturbative framework, is
also necessary  that, after our manipulations, the free actions of the
composites  
emerge in the effective action. To this aim  we  make use of  the
arbitrariness in the definition of 
the regularized action and perform a suitable choice of irrelevant
terms which help to construct the effective action of the composites.

In our previous work~\cite{Cara,CPS1}  we got a proof of the
spontaneous breaking of the chiral 
symmetry in QCD in the framework of our perturbative approach.
Exploratory applications of this approach have been performed in the
study of the 
pion-nucleon interaction~\cite{DeFr2} and the high temperature QCD
phase transition~\cite{DeFr3}. 
These works are limited by the fact that in the broken phase the
chiral symmetry is not realized nonlinearly. 
This property has been included in~\cite{CPS1} where we proved that,
in the absence of an explicit breaking due to the 
regularization, our approach generates the usual expansion of the
chiral models in momenta and masses. 

In the present paper we evaluate the amplitude of the decay of the $\pi_0$
into two photons and show that our effective action correctly
reproduces the chiral anomaly. These results are {\em per se}  interesting,
but they have in the present context a special relevance, related to
the fact that 
in our effective action there is an explicit breaking of the chiral symmetry
due to the regularization. This is 
because we are at the moment unable to treat the gluon composites analytically
in analogy to the quark composites and therefore, to have access to
the non-perturbative (in the gauge coupling constant) regime of the
gluon field  we  
are forced to define the theory on a lattice (but we emphasize that
our approach is  
otherwise general). It is likely that in the near future we will have a 
chirally invariant form of this regularization also in the presence of
non-Abelian fields~\cite{GW,Neu,Lusher}. But for the time being to get rid of
the spurious  states of the fermions on a lattice we must introduce the so
called Wilson term ~\cite{Wilson}, which explicitly  breaks the chiral
invariance of our effective action. It is therefore very important  to
ascertain 
whether the Wilson parameter can be taken  arbitrarily small in order to  avoid
this unpleasant consequence, but at the same time  reproducing the chiral
anomaly  of QCD. 

This problem is made somewhat more intriguing by the fact that in our approach
the quarks are perturbatively confined, because in the broken vacuum
they acquire a large mass which does not allow their propagation to
any finite order 
of our perturbative expansion. But if the quarks do not have any
poles, why should we 
worry about the spurious ones and introduce the Wilson term? At the
same time, if we 
omit this term, how can we reproduce the chiral anomaly?

The present findings  provide an answer to these questions.
We find the standard results subject to a condition on the  quark
effective mass which is 
naturally satisfied in our approach~\cite{CPS2}. Then  we can assume
the Wilson parameter $r$ 
arbitrarily small and forget the Wilson term in  
our expansion in the strong sector, even though we
must retain it in the electromagnetic amplitude which is non analytic
in this  parameter. 

In the next Section, for the convenience of the reader, we summarize
the essential  
steps of the derivation of the effective action. In Section 3 we
evaluate the electromagnetic decay amplitude of the $\pi_0$, in Section 4
we evaluate the anomaly and in Section
5 we present our conclusions.

\section{The quark-composites approach}

 We assume the modified partition function 
\begin{equation}
Z = \int [dV]  [d\overline{\lambda} d\lambda]\; \exp[- S_{G} -S_{Q} -S_C], 
\end{equation}
where $S_{G}$ is the  action of the gauge fields, that is the
Yang-Mills and Maxwell actions,  $S_Q$ is the action of the
quark  fields  
and $S_C$ {\it is a four fermions irrelevant operator} which provides
{\em the kinetic terms  }
for the quark composites with the quantum numbers of the chiral
mesons. Therefore it will  
not have the form of the Nambu-Jona-Lasinio~\cite{NJL} or
Gross-Neveu~\cite{GN} models, 
that is of the so-called chirally extended QCD or $\chi$QCD (see also
for example~\cite{Kogut:1997}). $\lambda$ is the quark field while the
gluon and electromagnetic  
fields are  associated to the link variables $V_{\mu}$. Differentials
in square brackets  
are understood to be the product of the differentials  over the
lattice sites and  
the internal indices. All the fields live in an euclidian lattice of
spacing $a$.   

We introduce the following notation for the  sum over the lattice 
\be
(f,g)  =   a^4 \sum_x f(x) g(x).
\ee
In this notation the quark action is
\be
S_Q= (\overline{\lambda}, Q \lambda) + m_q (\overline{\lambda}, \lambda ).
\ee
As already stated we will use the Wilson form of the quark wave operator
\be
Q  =   \gamma_\mu  \, \overline{\nabla}_{\mu} - a \, {r \over 2} \, \Box.
\ee
The symmetric derivative $ \overline{\nabla}_{\mu}$ and the Laplacian
$ \Box$ are  
covariant and are defined in terms of the right/left derivatives
\be
( \nabla^{\pm}_{\mu})_{x \, y}  =   \pm {1\over a} \left( \delta_{x
    \pm \hat{\mu},\,y}  V_{ \pm \mu} (x)
  -  \, \delta_{x \, y} \right) 
\ee
according to
\begin{eqnarray}
\overline{\nabla}_{\mu} & = &{ 1 \over 2} \left( \nabla^{+}_{\mu} +
  \nabla^{-}_{\mu} \right) \\
\Box & = & \sum_\mu  \nabla^{+}_{\mu} \,  \nabla^{-}_{\mu} =
           \sum_\mu  \nabla^{-}_{\mu} \,  \nabla^{+}_{\mu} =
           \sum_\mu  {1\over a} \left(  \nabla^{+}_{\mu} -
  \nabla^{-}_{\mu} \right)
\end{eqnarray}
We adopt the standard conventions 
\begin{eqnarray}
 V_{\mu}(x) &=& \exp \left[ E \, a\, A_{\mu}^{em} \left( x + {1 \over
 2} \,\hat{\mu}  \right) +  g \, a\, A_{\mu}^{YM}  \left( x + {1 \over
 2} \,\hat{\mu}  \right)\right] \\
 V_{-\mu}(x) &=&  V^\dagger_{\mu}(x-\hat{\mu}),
\end{eqnarray}
where $E$ is a charge matrix.

The chiral composites are the pions and the sigma
\be 
  \vec{\hat{\pi}}  =  i\,k_{\pi}\,a^{2}\,\overline{\lambda} \gamma_5
  \vec{\tau}  \lambda,\;\;\;
  \hat{\sigma}  =  k_{\pi}\,a^{2}\,\overline{\lambda} \lambda.    
\ee
 $\gamma_5$ is assumed hermitian, the $\vec{\tau}$'s are the Pauli matrices and a factor of dimension
(length)${}^2$, necessary to give the composites the dimension of a
scalar, has been written in the form $a^2 k_{\pi}$.

Since for massless quarks the QCD action is chirally invariant, the
action of the chiral mesons must be, 
apart from a linear breaking term, $O(4)$ invariant. It must then have the form
\be
S_C=  { 1\over 4}\< (\hat{\Sigma}^\dagger,C \hat{\Sigma}) \> - {
  1\over 4\, a^2} 
 \<
(\chi^\dagger, \hat{\Sigma}) +( \hat{\Sigma}^\dagger, \chi)\>,\label{actionC2} 
\ee
where
\be
\hat{\Sigma}= \hat{\sigma} - i \;\vec{\tau} \cdot
\vec{\hat{\pi}},\;\;\; \chi = s - i \;\vec{\tau}  \cdot
\vec{p},\;\;\; \< A \> = \hbox{tr}^{isospin} A.
\ee
 We introduced the sources $s$ and $\vec{p}$ of the sigma and pion
 fields (their coupling to the quarks differ by a factor 
$k_{\pi}$ from  the notation of~\cite{GL}). 

Heuristic considerations, based on experience with simple, solvable models,
lead~\cite{Cara} to the following form of the wave operator of the 
chiral composites 
\be
 C = -{\rho^4 \over a^4} \; { 1 \over - \Box + \rho^2/a^2} \;, 
\ee
where $\rho$ is a dimensionless parameter. The irrelevance by power counting of
$S_C$ requires that in the continuum limit $\rho$ do not to vanish and $k_{\pi}$, as well as the product
$k_{\pi} \rho$, do not diverge.  Under these conditions operators of dimension higher than 4 are
accompanied by the appropriate powers of the cut-off. The operator $C$ behaves, as a function of the
distance, as the propagator of a particle with a mass divergent at least as the cutoff. Therefore, 
even though strictly speaking it is nonlocal, its departure from locality is very mild
in general, and very small with our present choice of $\rho \sim 1/a$ which makes the 
mass appearing in $C$ divergent as the square of the cutoff. Ultimately,
however, its irrelevance can be proven by showing that its local
approximation $C \sim -(\rho^2 / a^2) - \Box $  yields the same
results (at the cost of more involved calculations).
 This proof is at present not complete.

We replace the chiral composites by the auxiliary fields
\be
\Sigma = \Sigma_0 - i \vec{\tau} \cdot \vec{\Sigma}  
\ee
 by means of the 
Stratonovich-Hubbard transformation~\cite{Stra}. Ignoring, as we will
systematically do in the sequel, 
field independent factors, the partition function can be written 
\begin{eqnarray} 
   Z  & =  &\int [dV] [d \bar{\lambda} d\lambda]
   \left[{d\Sigma\over\sqrt{2 \pi}}\right]   \exp\left[-S_{G} -  S_0
 + \left(\overline{\lambda},(D - Q)\lambda\right)   \right] \nonumber \\
 & = &
 \int [dV]  \left[{d{\Sigma}\over\sqrt{2 \pi}}\right]   
   \exp\left[-S_{G} -  S_0  +
   \mbox{Tr} \ln ( D - Q ) \right],  \label{mass} 
\end{eqnarray}
where ``Tr'' is the trace over both space and internal degrees
of freedom and
we introduced  the functions of the auxiliary fields
\begin{eqnarray}
S_0  &=& -  {
   1\over 4} \rho^4 \< (\Sigma^\dagger,(a^4 C)^{-1} \Sigma) \>  ,  \label{schi}
\\ D &=& {1-\gamma_5 \over 2}\, k_{\pi}\,\left[\rho^2
   \Sigma +  \chi\right]     + {1+\gamma_5 \over 2}\, 
 k_{\pi}\,\left[\rho^2 \Sigma^\dagger + 
   \chi^\dagger\right] .
\end{eqnarray}
Eventually we will set $s=m$, the breaking parameter of the chiral
symmetry, which must be distinguished from the quark mass $m_q$. The
scaling with the lattice spacing that we will assume for $m$ will render
the corresponding term irrelevant. The quark mass is absorbed in the breaking
parameter $m$ according to  
\be
 m \rightarrow m - { 1 \over k_{\pi}} m_q.
\ee
Since $1/ k_{\pi}$ will play the role of an expansion parameter, the
contribution from $m_q$ will be sub-leading in our expansion.

The derivative nature of the couplings of the pions~\cite{Wein} is exhibited 
after the transformation
\be
\Sigma  =  R\,  {U}    \label{tras}
\ee
where
$U$ is an element of SU(2), and
\be
 R^2 = \Sigma_0^2 + \vec{\Sigma}^2.
\ee
The volume element
\be
\left[{d\Sigma\over\sqrt{2 \pi}}\right] = \left[{d\,R\over\sqrt{2
      \pi}}\right]  [d\, U] \exp \sum_x 3 \ln \, R 
\ee
provides the Haar measure $ [d \, U]$ over the group.
We get the effective action  
\begin{eqnarray}
\tilde{S} &=&  {1\over 4}\sum_{\mu} \<(\nabla_{\mu} (R \,{U}^\dagger),
 \nabla^{\mu} (R \,{U}) ) \> 
 + { \rho^2 \over 2 a^2} (R,R)  - \mbox{Tr } \ln  R  - \sum_x 3 \, \ln \, R
\nonumber\\
& &  -{1\over 2} \, \mbox{Tr } \ln\left(1 - {1\over
 \rho^2\, R}\, 
 \chi^\dagger {U} \right)    - {1\over 2} \,\mbox{Tr } \ln\left(1 + 
 {1\over \rho^2\, R}\,  {U}^\dagger \chi
 \right)\nonumber  \\
& & - \mbox{Tr } \ln\, \left( 1 + D^{-1} Q \right) \,,
\end{eqnarray}
where $\nabla_{\mu}$  is the right or left covariant
derivative.

After our manipulations the partition function becomes
\be
   Z   =  \int [dV] \left[{d R \over\sqrt{2 \pi}}\right]  
   \left[ d {U} \right]   \exp\left[-S_{G} -\tilde{S}  \right].  
\ee
The  minimum of $\tilde{S}$ is obtained for  
\be
U = 1,\;\;\;\overline{R}= \sqrt{\Omega}
{ 1 \over a \rho}\left[ 1 -  { a m \over 2 \rho \,\sqrt{\Omega}} \right] ,
\ee
where $\Omega$ is the total number of quark components. In our case,
by collecting the spinorial, colour and flavour indices we get
\be
\Omega = 24.
\ee
The expansion around the minimum is naturally organized as a series in
$1/\sqrt{\Omega}$. In this framework we have
a realization of the spontaneous breaking of the chiral invariance in QCD.

The dominant part of the Lagrangian density, neglecting the
fluctuations of $R$ and 
terms arising from the expansion of $\overline{R}$ with respect to $m$
is identical 
to that of the chiral models~\cite{Wein,GL}
\be
{\cal L}_2 =  {1\over 4}  f_{\pi}^2  \< \nabla_{\mu} {U}^\dagger
\nabla^{\mu} {U}      
  - 2 \, B  \left( \chi^\dagger {U}  +  {U}^\dagger \chi  \right) \>, 
\ee
with the identifications
 \be
 f_\pi = { \sqrt{ \Omega }\over a \rho},\;\;\;
B= { 1\over 2 \, a^2 f_{\pi}}.
 \ee
If we confine ourselves to this leading term, we must assume $ f_\pi =
92\;\hbox{Mev}$.
After these positions we recognize that the expansion in $1/\sqrt{\Omega}$ 
is equivalent to that one in $1/f_\pi$ .
Note that the above definition implies that $\rho \sim 1/a$. Other
scalings with the lattice spacing are possible, but will not be
considered here. 

If we introduce the  pion field $\pi$ according to 
\be
U =  \exp \left({i\over f_\pi} \vec{\tau}\cdot\vec{\pi} \right), 
\ee 
we have for the pion mass $m_{\pi}$ and the chiral condensate the
relations
\be
 m_{\pi}^2 = 2mB,\;\;\; k_{\pi}\<0|\overline{\lambda} \lambda |0\> = 
 2 \, f_{\pi}^2 B. 
\ee
The presence of the factor $k_{\pi}$ is due to the fact that the source
$s$ has a coupling to the quark fields that differs by this factor
from the conventions 
of~\cite{GL}. It should also be noted that in the present case $m$ vanishes
while $B$ diverges in the continuum limit.      

Let us examine the mass of the quarks and the $\sigma$ in the broken vacuum. 
According to eq.~\reff{mass} the quark effective mass 
\be
M_Q=k_{\pi}\,\rho^2\, \bar{R} = k_{\pi}\,\rho^2\, f_\pi,
\ee 
is $O(k_{\pi} f_{\pi})$, and therefore the quarks are perturbatively confined.
Whether their mass is or is not divergent 
in the continuum limit, depends on how the product $k_{\pi} \rho$
scales with the  
lattice spacing. The $\sigma$ instead has a  mass $\sqrt{2} \rho /a$ which 
is always divergent in the continuum limit.

\section{The $\pi_0 \rightarrow \gamma \gamma$ decay}

In this Section all the functions are in momentum space. We will perform
the calculations to leading order in all the couplings, namely to order $1 / f_{\pi}$,
$ e^2 /( 2 \pi)^2$, while the Yang-Mills fields will be suppressed. The
gauge fields $A$ therefore, are only photon fields.

The  amputated amplitude  ${\cal I}_{\alpha\beta}$ for the decay of the $\pi_0$ into two photons
is related to the three-point function $ \< \pi_0(q)
A_{\alpha}(k_1)A_{\beta}(k_2)\>$ according to 
\begin{eqnarray}
\lefteqn{  \< \pi_0(q)  A_{\alpha}(k_1)A_{\beta}(k_2)\> =
  } \\
& & = {\cal I}_{\alpha\beta}(k_1,k_2) \,
G_{\pi}(q^2) \, G_A(k_1^2) \, G_A(k_2^2) \, (2 \pi)^4 \,
  \delta^4 (q + k_1 + k_2), \nonumber
\end{eqnarray}
where the $G$ are the free propagators of the
pion and the photons. ${\cal I}_{\alpha\beta}$ can be obtained by taking functional derivatives of the
effective action with respect to the Maxwell and the pion fields   
\begin{eqnarray}
\lefteqn{ (2 \pi)^4 \, \delta^4 (q + k_1 + k_2)\,{\cal
  I}_{\alpha\beta}(k_1,k_2)   = } \\  \label{icall} 
 & = & \left. {\delta^2 \over \delta A_\alpha(k_1) \delta
 A_\beta (k_2)}\, \hbox{Tr } 
\left[ {1\over  D-Q }\,{\delta D \over \delta \pi_3(q) }  \right]
\right|_{\vec{\pi}=A=0}
\nonumber\\
 & = &  
\left. {\delta^2 \over \delta A_\alpha(k_1) \delta
 A_\beta (k_2)}\, \hbox{Tr } \left[ 
\left( 1 +   {1\over  M_Q - Q }\,  Q 
  \right) \,{1\over M_Q}\,{\delta D   \over \delta  \pi_3(q) }  \right] 
\right|_{\vec{\pi}=A=0}  \nonumber\\
 & = &  
\left. {\delta^2 \over \delta A_\alpha(k_1) \delta
 A_\beta (k_2)}\, \hbox{Tr } \, \left[
  {1\over  M_Q - Q }\, {1\over 2} \left\{ Q\, ,\,{1\over M_Q}{\delta D
  \over \delta  \pi_3(q) }  \right\} \right]  
\right|_{\vec{\pi}=A=0}.  \nonumber
\end{eqnarray}
It is easy to relate the above expression the anomalous
triangle graph identifying vertices and propagators. Let
\be
  \left. \left({1\over M_Q - Q
        }\right)(k_1,k_2)\,\right|_{\vec{\pi}=A=0} = \delta_{k_1,k_2}
  \, S(k_1). 
\ee
$S$ is the free propagator of the quark in the broken vacuum
\be
S(k) =
\left( - i \, \bar{\kslash} + W(k) \right)^{-1} 
\ee
with
\begin{eqnarray}
W(k )& = & a \,{r\over 2} \hat{k}^2 +  M_Q \\
\bar{k}_\alpha &  = & { 1 \over a} \,\sin a \, k_\alpha \\
\hat{k}_\alpha & = & { 1 \over a} \,2 \sin a \,{k_\alpha \over 2}.
\end{eqnarray}
Having in mind the continuum limit, we neglected $\sqrt{\Omega}
k_{\pi}  m$ with respect to $M_Q$. 

Next we identify the vertices. The electromagnetic vertex is 
\be
\left.{\delta Q (k_1,k_2)\over \delta A_\alpha(k) }   \,
\right|_{\vec{\pi}=A=0} =  E\,   V_\alpha(k_1,k_2)\, 
(2 \pi)^4 \delta^4(-k_1+k_2+k) 
\ee
with
\be
 V_\alpha(k_1,k_2)  =   V_\alpha\left({k_1+k_2\over 2}\right)
\ee
and 
\be
V_\alpha(k) = {\partial \over \partial k_\alpha}\, S^{-1}(k). \label{Ward}
\ee
The quark-charge matrix $E$ is  defined as
\be
 E = {e\over 6} (1 + 3 \tau_3)
\ee
where $e$ is the electric charge of the proton.

The anomalous vertex is
\be
{1\over 2} \left\{ Q, { 1 \over M_Q} \left.{\partial D(k_1,k_2) \over \partial
     \pi_3(q)} \right\} \right|_{\vec{\pi}=A=0} =  
i{ 1 \over f_{\pi}}\,\gamma_5  \tau_3 W(q) (2 \pi)^4 \, \delta^4(-k_1
+ k_2 + q) . 
\ee
Using the above equations the decay amplitude becomes
\begin{eqnarray}
{\cal I}_{\alpha\beta}(k_1,k_2) &=&  i {1\over  f_\pi} \int \left({dk\over
    2 \pi}\right)^4 \hbox{tr } \left[ E^2 \tau_3 \, \gamma_5 W(k)
  S(k+k_1) \right.
\nonumber\\
   & & \left. V_\alpha(k+k_1,k) S(k) V_\beta(k, k-k_2)S(k-k_2)\right],
\end{eqnarray}
and after the sum over colour and isospin indices 
\begin{eqnarray}
\lefteqn{{\cal I}_{\alpha\beta}(k_1,k_2)  = - 2 \,i {e^2\over f_\pi}
  \int \left({dk\over 2 \pi}\right)^4   W(k) } 
\nonumber \\
&& \hbox{tr}^{spin}\left[
\gamma_5 S(k+k_1) V_\alpha(k+k_1,k) S(k) V_\beta(k, k-k_2)
  S(k-k_2)\right]. \label{I}
\end{eqnarray}

Let us develop the
expression~\reff{I} in series of the photon momenta $k_1$ and $k_2$.
By using the Ward identity~\reff{Ward} in the form
\be
  S \, V_\alpha \, S = S \, \partial_\alpha S^{-1} \, S = -
   \partial_\alpha S, \label{id}
\ee
where all the functions are evaluated at the same momentum $k$, one
can reduce the number of $\gamma$-matrices appearing in the trace. It is
easy to show in this way that the first non-vanishing contribution comes 
at order $k_1 k_2$ and is given by 
\begin{eqnarray}
{\cal I}_{\alpha\beta}(k_1,k_2) & =  &  \,i \,{e^2\over f_\pi}
  \,k_{1\mu} k_{2\nu}\,
  \int \left({dk\over 2 \pi}\right)^4  \hbox{tr}^{spin} \left\{
  \vphantom{1\over 2} \gamma_5 W
  \right.\nonumber  \\ 
& & \left[ {1\over 2}   S \partial_\mu
  V_\alpha S 
  \partial_\nu V_\beta S  + 2    \partial_\mu S  V_\alpha S 
  V_\beta \partial_\nu S \right.  
\nonumber\\
& & \left. \left.  \vphantom{1\over 2}+  \partial_\mu S  V_\alpha S
  \partial_\nu V_\beta S  +    S  \partial_\mu V_\alpha S 
   V_\beta \partial_\nu S \right] \right\}  
\end{eqnarray}
which, after a few integrations by parts, becomes
\begin{eqnarray}
{\cal I}_{\alpha\beta}(k_1,k_2) & =  & -  \,i  \, {e^2\over  f_\pi}
  \,k_{1\mu} k_{2\nu}\,
  \int \left({dk\over 2 \pi}\right)^4  \hbox{tr}^{spin}  \left\{ \gamma_5 W
  \right. \nonumber \\ 
& & \left. \left[ \partial_\mu  S V_\alpha \partial_\nu S 
  V_\beta S  +      S  V_\alpha \partial_\mu S 
  V_\beta \partial_\nu S \right]\right\}.
\end{eqnarray}
By using the explicit expressions of $S$ and $V$ we get 
\begin{eqnarray}
{\cal I}_{\alpha\beta}(k_1,k_2) & =  & - 2 \,i \,{e^2  \over f_\pi}
  \,k_{1\mu} k_{2\nu}\,
  \int \left({dt\over 2 \pi}\right)^4  \hbox{tr}^{spin} [ \gamma_5
  \gamma_\mu \gamma_\alpha \gamma_\nu \gamma_\beta ] {W \over d^3} \nonumber
  \\ 
& &   \cos t_\alpha \cos t_\nu \cos t_\beta \left[W \cos
  t_\mu - 4   \partial_\mu W \sin  t_\mu\right] 
\end{eqnarray}
where
\be
d = \bar{t}^2 + W^2(t).
\ee
We assume, as usual, that in the continuum limit
\be
 \lim_{a\to 0}   {a M_Q\over r} = 0  \; , \label{condition}
\ee
a condition which can be naturally satisfied in our approach~\cite{CPS2}.
Then we see that the integral takes its contribution only from the
pole and gives the 
well known result~\cite{KS}   
\begin{eqnarray}
{\cal I}_{\alpha\beta}(k_1,k_2) & =  & - 8 \,i {e^2 \over f_\pi}
  \, \epsilon_{\mu\nu\alpha\beta} k_{1\mu} k_{2\nu}\,
  \int \left({dt\over 2 \pi}\right)^4  \cos t_2 \cos t_3 \cos t_4
  \partial_1 {\sin t_1 \over d^2}\nonumber  \\
& = &  -  \,i {1 \, \over f_\pi} \left({e\over 2 \pi}\right)^2
  \, \epsilon_{\mu\nu\alpha\beta} k_{1\mu}  k_{2\nu}. \label{result}
\end{eqnarray}

\section{The chiral anomaly}

Apart from the explicit breakings, that is the
source  and the Wilson terms, our partition function  
is exactly invariant under the  symmetry transformations 
\be
U \to g_R U g_L^\dagger , \label{transformations}
\ee
with $(g_R,g_L)\in$ SU(2)$\times$SU(2). While the presence of the
source term is necessary to 
provide a mass term to the pions, the Wilson term is a residue of the doubling
problem of lattice fermions which  induces a departure from the
general structure of the chiral Lagrangians. Nonetheless this term is
responsible for the correct chiral anomaly in lattice QCD, which,
as it is well known, is deeply related to the 
electromagnetic decay of the $\pi_0$. 


Let us derive the general Ward identities. For the sake of simplicity
hereafter we shall restrict the field $R$ at his extremal value and we
neglect its fluctuations.

For
an arbitrary infinitesimal transformation 
\be
U  \to U + \delta U
\ee
with
\be
\delta U = w_R \, U - U  w_L.
\ee
The variation of the first term of $\tilde{S}$ is
\begin{eqnarray}
\lefteqn{{1\over 2}\, \delta\, \hbox{Tr } \sum_\mu \left[
  \nabla_\mu^+ U^\dagger 
  \nabla_\mu^+   U \right]\, = } \\
& = &  - \delta \, \hbox{Tr } \sum_\mu \left[ U_{x+\mu}^\dagger U_{x} +
  U_{x}^\dagger  U_{x+\mu} \right] \nonumber \\
& = & -  \hbox{Tr } \sum_\mu \left[-   U_{x+\mu}^\dagger  \delta
 U_{x+\mu}  U_{x+\mu}^\dagger  U_{x} +
 U_{x+\mu}^\dagger\delta U_{x} -   U_{x}^\dagger  \delta
 U_{x}  U_{x}^\dagger  U_{x+\mu} +
  U_{x}^\dagger \delta  U_{x+\mu} \right] \nonumber \\
& = & -  \hbox{Tr } \sum_\mu \left[ - U_{x}  U_{x+\mu}^\dagger   +
U_{x+\mu}  U_{x}^\dagger   \right] \left[ \delta
U_{x+\mu}  U_{x+\mu}^\dagger -  \delta U_{x}  U_{x}^\dagger \right]
 \nonumber \\
& = & -  \hbox{Tr } \sum_\mu \left[ - U_{x}  U_{x+\mu}^\dagger   +
 U_{x+\mu}  U_{x}^\dagger   \right] \nabla_\mu^+  \left[  \delta
  U_{x}  U_{x}^\dagger \right] 
 \nonumber \\
& = & \hphantom{-}  \hbox{Tr } \sum_\mu \left\{ \nabla_\mu^- \left[ - U_{x}
  U_{x+\mu}^\dagger   + 
 U_{x+\mu}  U_{x}^\dagger   \right] \right\} \left[  \delta
  U_{x}  U_{x}^\dagger \right] 
 \nonumber \\
& = &  \hphantom{-} \hbox{Tr } \sum_\mu \left\{ \nabla_\mu^- \left[ - U
\left( \nabla_\mu^+  U^\dagger \right)   + 
\left( \nabla_\mu^+  U  \right) U^\dagger   \right] \right\} \left[  \delta
  U  U^\dagger \right] 
 \nonumber \\
& = &  \hphantom{-} \hbox{Tr } \sum_\mu \left\{ \nabla_\mu^- \left[ - U
\left( \nabla_\mu^+  U^\dagger \right)   + 
\left( \nabla_\mu^+  U  \right) U^\dagger   \right] \right\}
  \left[ w_R - U w_L U^\dagger \right] 
 \nonumber \\
& = &  \hphantom{-} \hbox{Tr } \sum_\mu \left\{ \nabla_\mu^- \left[ - U
\left( \nabla_\mu^+  U^\dagger \right)   + 
\left( \nabla_\mu^+  U  \right) U^\dagger   \right] \right\}
   w_R  \nonumber \\
&   &  + \hbox{Tr } \sum_\mu \left\{ \nabla_\mu^- \left[ - U^\dagger
\left( \nabla_\mu^+  U \right)   + 
\left( \nabla_\mu^+  U^\dagger  \right) U   \right] \right\}
   w_L  \nonumber
\end{eqnarray}
which has the form of the divergence of a current.
More precisely for a vectorial transformation, that is when
\begin{eqnarray}
                         g_L & = & g_R \\
              w_L & = & w_R \, = \,  w_V \label{vector}
\end{eqnarray}
we get the vector current
\be
\vec{\cal{V}}_\mu =  {1\over 2} \left\< \vec{\tau}
   \left\{ \left[  \nabla_\mu^+  U , U^\dagger \right] 
      +  \left[  \nabla_\mu^+  U^\dagger , U \right] \right\}  \right\>.
\ee
While for an axial transformation 
\begin{eqnarray}
                         g_L & = & g_R^\dagger \\
              w_L & = & - w_R \, = \, -  w_A \label{axial}
\end{eqnarray}
we get
\be
\vec{\cal{A}}_\mu =  {1\over 2}  \left\<  \vec{\tau}
   \left[ \left\{  \nabla_\mu^+  U , U^\dagger \right\} 
      -  \left\{  \nabla_\mu^+  U^\dagger , U \right\} \right]\right\>.
\ee
Let us consider now the explicit breaking at linear level
\be 
\delta\, \hbox{Tr } \left( \chi U^\dagger + U \chi^\dagger \right) 
 =  
 \hbox{Tr } \left[  \left( U \chi^\dagger - \chi U^\dagger \right) w_R -
                  \left( \chi^\dagger U - U^\dagger \chi \right) w_L
 \right].
\ee
From this expression in the case of a vector transformation  we get
\be 
\delta\, \hbox{Tr } \left( \chi U^\dagger + U \chi^\dagger \right) 
  =  
 \hbox{Tr } \left\{ \left[U,\chi^\dagger \right] + \left[U^\dagger,
     \chi \right] \right\} w_V 
\ee
while for an axial transformation
\be 
\delta\, \hbox{Tr } \left( \chi U^\dagger + U \chi^\dagger \right) 
  =  
 \hbox{Tr } \left[ \left\{U,\chi^\dagger \right\}  -  \left\{U^\dagger,
     \chi \right\} \right] w_A.
\ee
From these expressions we get, in the absence of the fermionic
determinant, when $\chi= m$
\begin{eqnarray}
 \sum_\mu  \nabla_\mu^-\, \vec{\cal{V}}_\mu  &=& 0 \\
 \sum_\mu  \nabla_\mu^-\, \vec{\cal{A}}_\mu  &=& 
m_\pi^2 \,\left \< \vec{\tau} \left( U - U^\dagger \right) \right\>
 \label{zero} 
\end{eqnarray}
which is nothing but the classical equation of motion.

Let us now evaluate the variation of the fermionic determinant. Since 
we have already taken the linear contribution of the breaking term we
will set now for simplicity $\chi=0$. Then
\be
 D  = M_Q \left[ {1-\gamma_5 \over 2}\, U +  {1+\gamma_5 \over 2}\,
   U^\dagger \right] 
\ee
and its variation is
\begin{eqnarray}
\delta  D  &=&  M_Q \left[ {1-\gamma_5 \over 2}\, \delta U +  {1+\gamma_5
    \over 2}\,   \delta U^\dagger\right]\\
 &=&  M_Q \left[ {1-\gamma_5 \over 2}\, \delta U -  {1+\gamma_5
    \over 2}\,   U^\dagger \delta U  U^\dagger \right].  \nonumber 
\end{eqnarray}
Therefore
\begin{eqnarray}
\lefteqn{
\delta \hbox{Tr } \ln ( D - Q ) \,  =  \,  \hbox{Tr } (D-Q)^{-1} \delta D
\, =}\\
& = &   \hbox{Tr } (D-Q)^{-1} M_Q \left[ {1-\gamma_5 \over 2}\, \delta
  U -  {1+\gamma_5 
    \over 2}\,   U^\dagger \delta U  U^\dagger \right]  \nonumber \\
& = & \hbox{Tr } M_Q \left[- (D-Q)^{-1} {1-\gamma_5 \over 2}\,  U +
  {1+\gamma_5  
    \over 2}\,   U^\dagger  (D-Q)^{-1} \right] w_L +  \nonumber \\
&   &  \hbox{Tr } M_Q \left[ {1-\gamma_5 \over 2}\,  U  (D-Q)^{-1} -
   (D-Q)^{-1} {1+\gamma_5  \over 2}\,   U^\dagger  \right] w_R.   \nonumber 
\end{eqnarray}
If we specialize to a vector transformation 
\begin{eqnarray}
\delta \hbox{Tr } \ln ( D - Q ) 
& = & -\, M_Q\, \hbox{Tr } \left[ \, (D-Q)^{-1} , {1-\gamma_5 \over
      2}\,  U + 
  {1+\gamma_5  
    \over 2}\,   U^\dagger   \right] w_V   \nonumber \\
& = & -\, \hbox{Tr } \left[ \, (D-Q)^{-1} ,  D   \right] w_V
\nonumber \\ 
& = & -\, \hbox{Tr } \left[ \, (D-Q)^{-1} ,  Q   \right] w_V
\nonumber \\ 
& = & -\, \hbox{Tr } (D-Q)^{-1}   \left[ \, Q ,   w_V \right]. \label{V}
\end{eqnarray}
While for an axial transformation 
\begin{eqnarray}
\delta \hbox{Tr } \ln ( D - Q )   
& = & M_Q\, \hbox{Tr } \left\{ \, (D-Q)^{-1} , {1-\gamma_5 \over
      2}\,  U - 
  {1+\gamma_5  
    \over 2}\,   U^\dagger   \right\} w_A   \nonumber \\
& = & -\, \hbox{Tr } \left\{ \, (D-Q)^{-1} , \gamma_5 D   \right\} w_A
\nonumber \\ 
& = & -\, \hbox{Tr }\gamma_5  \left\{ \, (D-Q)^{-1} , D   \right\} w_A
\nonumber \\ 
& = & - \, \hbox{Tr }\gamma_5  \left\{ \, (D-Q)^{-1} , Q  \right\} w_A
\nonumber \\
& = &  - \, \hbox{Tr }  (D-Q)^{-1} \left(  Q \gamma_5  w_A +  w_A
    \gamma_5 Q    \right)  \nonumber \\
& = &  - \, \hbox{Tr }   (D-Q)^{-1} \left(  - \gamma_5 \left[
   Q,   w_A \right] 
 + \left\{ Q, \gamma_5\right\}  w_A  \right)  \label{A}
\end{eqnarray}
We arrive at the equations
\begin{eqnarray}
\sum_\mu  \nabla_\mu^-\, \vec{\cal{V}}_\mu
 &=& {2 \over f_\pi^2  }\, 
 \hbox{Tr } \vec{\tau}\, \left[ \, Q,  (D-Q)^{-1}    \, \right] \\
\sum_\mu  \nabla_\mu^-\, \vec{\cal{A}}_\mu  &=& 
m_\pi^2 \,\left \< \vec{\tau} \left( U - U^\dagger \right) \right\> -
{2 \over f_\pi^2  }\, \hbox{Tr } \vec{\tau}\, \left[ \,\gamma_5  Q,
 (D-Q)^{-1}    \, 
 \right] \nonumber  \\
& &  - \,{2 \over f_\pi^2  }\, \hbox{Tr }  \vec{\tau}\,  (D-Q)^{-1}
 \left\{ Q, \gamma_5\right\}.
\end{eqnarray}
where the new terms with respect to~\reff{zero} correspond to the fermionic
contributions to the currents. In particular the last term with the
anti-commutator originates the anomalous vertex which entered in the
computation of the decay rate of the $\pi_0$ of the previous section.
It is therefore this term which is responsible for the breaking, at
the quantum level, of the axial symmetry in the continuum limit.
The anomaly of the underlying QCD is correctly reproduced by our
effective chiral lagrangian.

\section{Conclusions}

We evaluated the decay amplitude of the electromagnetic decay of the
pion by an effective action derived from QCD in the quark composites
approach. This allowed us to treat in a unified way the anomalous 
and the electromagnetic vertices.

As a consequence of formula~\reff{result}, to leading order the decay
rate turns 
out to take the standard value of $7.63 \,\, \hbox{eV}$ (see for
example~\cite{IZ}), surprisingly
close to the 
experimental rate  $(7.37 \, \pm 1.5) \, \, \hbox{eV}$.
It would be interesting to check whether the strong corrections are
sufficiently small in our theory.

The present results allow us to establish a close relation between our
expansion for the chiral mesons and the chiral models. Since
the chiral anomaly is independent of the value of the Wilson parameter
$r$, provided it is 
different from zero, we can then assume $r=O(\Omega^{-n})$, with
$n$ arbitrarily large. Now there is no reason to believe that
the  amplitudes in the strong sector
are not analytic in $r$, and therefore studying the strong interactions in the 
framework of our
$1/\sqrt{\Omega} \sim 1/f_{\pi}$ expansion, we can
forget the Wilson term. We showed that in this case our theory, under
the standard  
condition~\reff{condition} which is naturally satisfied,
generates only terms of the chiral models~\cite{CPS1}. In the
electromagnetic sector, on the contrary, the amplitudes are not
analytic in $r$, as we have seen in the previous section, and the
Wilson term must be retained to get correct results.

We think that the quark-composites approach might prove useful  also
in numerical simulations. 
One can consider the action of eq.~\reff{mass} as an improved lattice
QCD action, where the chiral limit is obtained in the simple limit of 
zero breaking term, where the pion
mass vanishes (rather than by a fine tuning). There is a price to pay
because of the  
inclusion of the auxiliary fields, but this should be rewarded by a simpler 
evaluation of the quark determinant because of the dominance of the
diagonal contribution in configuration space.
There are indeed indications in this sense in the work presented
in~\cite{Brower:1995}.

\section{Acknowledgments}

It is a pleasure to thank Andrea Pelissetto for useful discussions.

\end{document}